\documentclass[letterpaper, 10 pt, conference]{ieeeconf}  

\IEEEoverridecommandlockouts                              

\usepackage{amssymb}
\usepackage[dvipsnames]{xcolor}
\usepackage{color, colortbl}
\usepackage{booktabs}
\usepackage{graphicx}
\usepackage{amsmath}
\usepackage{multirow}
\usepackage{caption}

\title{\LARGE \bf
VIMPPI: Enhancing Model Predictive Path Integral Control with Variational Integration for Underactuated Systems
}

\author{Igor Alentev$^{\;\dag\;1}$, Lev Kozlov$^{\;\dag\;1}$, Ivan Domrachev$^{\;\dag\;1, 2}$, Simeon Nedelchev$^{2}$, Jee-Hwan Ryu$^1$
\thanks{\dag Equal contribution. Alphabetic ordering.}%
\thanks{*This research has been financially supported by The Analytical Center for the Government of the Russian Federation (Agreement No. 70-2021-00143 dd. 01.11.2021, IGK
000000D730321P5Q0002)}
\thanks{$^{1}$ Interactive Robotic Systems Laboratory, Korea Advanced Institute of Science and Technology (KAIST)}%
\thanks{$^{2}$ Institute of Robotics and Computer Vision, Innopolis University, Innopolis, Russia.}%
}

\begin{document}
\definecolor{LightRed}{rgb}{1,0.7,0.7}

\maketitle
\thispagestyle{empty}
\pagestyle{empty}

\begin{abstract}
    This paper presents VIMPPI, a novel control approach for underactuated double pendulum systems developed for the AI Olympics competition. We enhance the Model Predictive Path Integral framework by incorporating variational integration techniques, enabling longer planning horizons without additional computational cost. Operating at 500-700 Hz with control interpolation and disturbance detection mechanisms, VIMPPI substantially outperforms both baseline methods and alternative MPPI implementations.
\end{abstract}

\section{Introduction}
The AI Olympics with RealAIGym~\cite{wiebe2022realaigym} competition challenges participants to develop controllers for complex robotic systems. This year's task focused on designing controllers for underactuated double pendulum systems — the pendubot and acrobot — with emphasis on maintaining the upper equilibrium position from various initial states \cite{10375556}.

Underactuated systems, characterized by fewer control inputs than degrees of freedom, present significant challenges in control engineering. The double pendulum, with its highly nonlinear and chaotic dynamics, serves as an excellent benchmark for evaluating control strategies. Controlling such systems requires balancing computational efficiency with planning accuracy, particularly when dealing with the complex dynamics that emerge from their underactuated nature.

Previous competitions \cite{10.24963/ijcai.2024/1043}\cite{wiebe2025reinforcementlearningrobustathletic} featured diverse learning-based approaches, including model-free reinforcement learning (AR-EAPO) \cite{ar_eapo}, Monte Carlo Probabilistic Inference for Learning Control (MC-PILCO) \cite{mcpilco}, and soft actor-critic (SAC) models with evolutionary fine-tuning \cite{evolsac}. While effective, these methods typically require extensive training data and carefully designed reward functions, limiting their practical applicability in scenarios where training resources are constrained or system dynamics change frequently.

Given these limitations, we explored sampling-based methods as an alternative approach. These methods operate directly on system dynamics without pre-training, making them particularly suitable for systems with well-understood physics. Recent advances in algorithms and parallel computing have made sampling-based methods increasingly viable for planning in highly dynamic environments and controlling systems with many degrees of freedom \cite{sampling_overview}. The double pendulum's chaotic yet well-modeled nature makes it an ideal candidate for such approaches.

Among sampling-based methods, Model Predictive Path Integral (MPPI) control has emerged as a powerful technique, with successful applications ranging from aggressive vehicle control \cite{williams2016aggressive} to various robotic systems. However, computational constraints typically limit the planning horizon in traditional MPPI implementations. Our approach addresses this fundamental limitation by incorporating a variational integrator into the rollout computations, enabling a substantially longer planning horizon without additional computational cost. To our knowledge, this is the first application of such an approach, and our controller significantly outperforms established benchmarks.

\section{Background}
Before presenting our approach, we first review the key concepts underlying our work: Model Predictive Path Integral control and variational integration techniques.

\subsection{Model Predictive Path Integral}
We model the double pendulum as a stochastic dynamical system:
\begin{equation}
    x_{i+1} = f(x_i, u_i + \delta u_i)
\end{equation}
where \(x_i \in \mathbb{R}^4\) represents the state vector of the pendulum at step \(i\), \(u_i \in \mathbb{R}\) denotes the control input (applied to either the first or second joint, depending on the configuration), and \(\delta u_i\) is a zero-mean Gaussian noise vector with variance \(\Sigma_u\), i.e., \(\delta u_i \sim \mathcal{N}(0, \Sigma_u)\), representing control input perturbations.

Given a time horizon \(t \in \{0, 1, 2, \dots, T-1\}\), the objective of the stochastic optimal control problem is to find a control sequence $u = (u_0, u_1, \dots, u_{T-1}) \in \mathbb{R}^{m \times T}$ that minimizes the expected value of a scalar cost-to-go function:
\begin{equation} \label{opt-problem}
    J = \min_u \mathbb{E}[S(x, u)],
\end{equation}

In our formulation, the cost-to-go function is defined as:
\begin{equation}
    S(x, u) = \phi(x_T) + \sum_{i=0}^{T-1} \Bigl( q(x_i, u_i, \delta u_i) + u_i^T R u_i \Bigr)
\end{equation}
where $q(x_i, u_i, \delta u_i)$ is the stage cost:
\begin{equation}
    q(x_i, u_i, \delta u_i) = \tilde{x}_i^T Q \tilde{x}_i + \gamma (u_i + \delta u_i)^T \Sigma_u^{-1} u_i
\end{equation}
Here, $\tilde{x} = x - x_{goal}$ represents the error between the current and desired state (which is always the upper equilibrium position), and the term $\gamma (u_i + \delta u_i)^T \Sigma_u^{-1} u_i$ penalizes outliers based on the distribution variance.

To solve this optimization problem (\ref{opt-problem}), \cite{williams_model_2017} proposes the following update law:
\begin{equation}
    u_i \leftarrow u_i + \frac{\sum_{k=1}^{K} \exp\Bigl(-\frac{1}{\lambda} \tilde{S}(\tau_{t,k})\Bigr) \, \delta u_{t,k}}{\sum_{k=1}^{K} \exp\Bigl(-\frac{1}{\lambda} \tilde{S}(\tau_{t,k})\Bigr)}
\end{equation}
where \(K\) is the number of random samples (rollouts), and \(\lambda \in \mathbb{R}^+\) is a hyperparameter (the inverse temperature). The control rollout is preserved between iterations with added noise, enabling precise convergence to optimal trajectories.

A key component is the rollout process which integrates the pendulum dynamics forward in time. Traditional Euler methods accumulate energy errors and require tiny timesteps (0.001s), severely limiting the planning horizon. While implicit integration and Runge-Kutta methods improve on this, they still have computational costs or energy drift issues. These limitations motivated our use of variational integrators for accurate simulation of underactuated system dynamics.

\subsection{Variational Integrator}

Variational integrators represent a fundamentally different approach to numerical integration compared to traditional methods. Rather than discretizing the equations of motion directly, these specialized symplectic integrators discretize the underlying variational principle (Hamilton's principle). This approach preserves the system's physical properties—such as energy and momentum—over long time periods, making it particularly valuable for simulating chaotic systems like the double pendulum.

In our implementation, we employ the Discrete Euler-Lagrange residual in Momentum form. This formulation allows us to work with momentum equations and accurately estimate the next configuration step in MPPI rollouts. Instead of solving a computationally expensive rootfinding problem at each integration step, we take an initial configuration guess and refine it using the discrepancies between estimated and ground truth momentum. The foundations of this approach were established by West~\cite{west_variational_integrators}.

The core of the variational integrator is the Discrete Lagrangian $\mathcal{L}_\mathrm{d}$, which approximates the action integral of the continuous Lagrangian $\mathcal{L}$ over a small time interval:

\begin{equation}
    \mathcal{L}_\mathrm{d}(q_n, q_{n+1}) \approx \int_{t_n}^{t_{n+1}} \mathcal{L}(q(t), \dot{q}(t)) \, dt
\end{equation}

For mechanical systems like our double pendulum, the Lagrangian is typically the difference between kinetic and potential energy: $\mathcal{L}(q, \dot{q}) = T(\dot{q}) - V(q)$. We implement a midpoint approximation of the Discrete Lagrangian:

\begin{equation}
    \mathcal{L}_\mathrm{d}(q_n, q_{n+1}) = \mathcal{L}\left(\frac{q_n + q_{n+1}}{2}, \frac{q_{n+1} - q_n}{\Delta t}\right)\Delta t
\end{equation}

This discretization preserves the symplectic structure of the system, ensuring that energy and momentum are conserved over long time periods—a crucial property for accurate simulation of the double pendulum dynamics.

The integration process begins with an initial guess for the next configuration, derived from an explicit Euler step:
\begin{equation}
    \hat{q}_{n+1} = q_{n} + v_{n} \Delta t
\end{equation}

We then refine this guess through an iterative process based on momentum matching. Since the true velocity is known, we can calculate the momentum as:
\begin{equation}
    p_{k} = M\left(q_{k}\right)v_k
\end{equation}

Alternatively, momentum can be estimated from two consecutive configurations using the discrete Lagrangian:
\begin{equation}
    \hat{p}_{k} = -\nabla_{q_{n+1}}\mathcal{L}_\mathrm{d}\left(q_{n}, q_{n+1}\right)
\end{equation}

The difference between these two momentum values guides our refinement process. We calculate the scaling factor for the configuration adjustment as the Jacobian of the gradient of the discrete Lagrangian with respect to the arguments:
\begin{equation}
    A = \nabla_{q_{k+1}}\left(\nabla_{q_{k}}\mathcal{L}_\mathrm{d}\right)
\end{equation}

The required configuration adjustment is then computed as the inverse of the scaling matrix multiplied by the momentum error:
\begin{equation}
    \Delta\hat{q}^i = A^{-1} \cdot \left(p_k - \hat{p}_k\right)
\end{equation}

Finally, we refine the next configuration approximation by subtracting this error:
\begin{equation}
    \hat{q}_{n+1}^{i + 1} = \hat{q}_{n+1}^{i} - \Delta \hat{q}^i
\end{equation}

By iterating this process, the initial configuration guess converges to an accurate solution that preserves the system's conservative properties. For our double pendulum system, we found that a single error correction iteration is typically sufficient, as subsequent error magnitudes become negligibly small. This makes the computational overhead of the variational integrator minimal compared to the benefits it provides in simulation accuracy and stability.

The key advantage of this approach is that it maintains the physical consistency of the simulation even with larger timesteps, which is crucial for extending the planning horizon in MPPI. This property forms the foundation of our Variational MPPI (VIMPPI) approach, which we describe in the following section.

\section{Method}
Building on the theoretical foundations described above, we now present our VIMPPI approach and its key components. Our method combines the sampling-based optimization of MPPI with the enhanced numerical stability of variational integration, supplemented by additional mechanisms to improve performance in practical scenarios. Together, these elements create a robust control framework capable of handling the challenging dynamics of underactuated pendulum systems.

\subsection{VIMPPI: Variational Integrator + MPPI}
The core innovation of our approach lies in replacing the standard explicit Euler integrator with a variational integrator in the MPPI rollout process. This enables us to use much larger timesteps without compromising stability. While traditional MPPI implementations are limited to timesteps of 0.001-0.005 seconds, our variational integrator allows timesteps of 0.02 seconds or larger while maintaining accuracy. This 4-20x increase in timestep size directly translates to a proportional extension of the planning horizon with the same computational budget.

This enhancement transformed our controller implementation. Rather than requiring a separate tracking controller as is common in MPPI applications, we implemented MPPI as a direct controller operating at 500-700 Hz. The resulting controller exhibits exceptional responsiveness, speed, and precision in both planning and execution, outperforming baseline controllers even before additional optimizations.

The ability to extend the planning horizon without increasing computational cost makes this approach valuable for resource-constrained platforms like small drones or embedded systems. By preserving the physical properties of the system even with larger timesteps, the variational integrator enables more accurate long-term predictions crucial for swing-up maneuvers and disturbance recovery.

\subsection{Control Interpolation}
While the variational integrator improved our planning capabilities, we encountered a challenge with high-frequency operation due to misalignment between the controller execution frequency and rollout timesteps. For example, with a 500 Hz controller (2ms period) but 20ms rollout timesteps, the remaining control sequence would become misaligned with the system state after applying the first control.

To address this, we implemented linear interpolation between control points to align the sequence with new timestep boundaries. This enhancement proved crucial in practice, significantly improving control precision and success rates in maintaining the pendulum at unstable equilibrium positions.

\subsection{Disturbance Detection and Warm Start}
To handle severe disturbances that could destabilize the pendulum, we developed a disturbance detection mechanism that monitors the system by comparing expected and actual state changes. When differences exceed a threshold, the system triggers a recovery process.

For efficient recovery, we implemented a warm-start mechanism using the AR-EAPO baseline controller to calculate an initial control sequence, which our MPPI algorithm then refines. This hybrid approach enables quick recovery from disturbances while maintaining the advantages of our variational integrator-based planning. Despite relying on another controller for initialization, our implementation significantly improved upon its results.

\section{Results}
To evaluate the performance of our VIMPPI controller, we conducted extensive experiments comparing it against various baseline methods. We implemented our controller using JAX \cite{jax2018github}, which enabled efficient parallelization across available hardware resources—a critical factor in achieving the computational performance needed for real-time control. The hyperparameters for the VIMPPI controller are shown in Table \ref{tab:mppi_parameters}.

\begin{table}
    \centering
    \caption{MPPI hyperparameters.}
    \label{tab:mppi_parameters}
    \begin{tabular}{ll}
        \hline
        \textbf{Parameter} & \textbf{Value}                               \\ \hline
        Horizon            & $20$                                         \\
        Samples            & $4096$                                       \\
        $\lambda$          & $50.0$                                       \\
        $\alpha$           & $1.0$                                        \\
        $\sigma$           & $0.2 \cdot I_2$                              \\
        $Q$                & $\text{diag}(10.0, 1.0, 0.10, 0.10)$         \\
        $R$                & $\text{diag}(0.10, 0.10)$                    \\
        $P$                & $10^6 \cdot \text{diag}(5.0, 5.0, 2.0, 2.0)$ \\
        Rollout $\Delta t$ & $0.02$ s                                     \\
        \hline
    \end{tabular}
    \vspace*{-5mm}
\end{table}

We compared our approach against several algorithms, including baselines from previous competitions (evolsac, History SAC, AR-EAPO, mcpilco, TVLQR). Moreover, to prove the unfeasibility of the approach using classical integration techniques, we implemented different integrators and tested against our proposed with the same setup horizon conditions: implicit (I), implicitfast (IF), explicit (E), and our Variational Integrator (VI). Testing was conducted on an NVIDIA RTX 4070 Ti Super with 20 test runs per controller using randomized disturbance patterns. This comprehensive evaluation framework allowed us to isolate the specific benefits of our variational integration approach while ensuring fair comparisons across methods.

\subsection{Pendubot}

\begin{figure}[t]
    \centering
    \begin{minipage}{\linewidth}
        \centering
        \includegraphics[width=1.\linewidth]{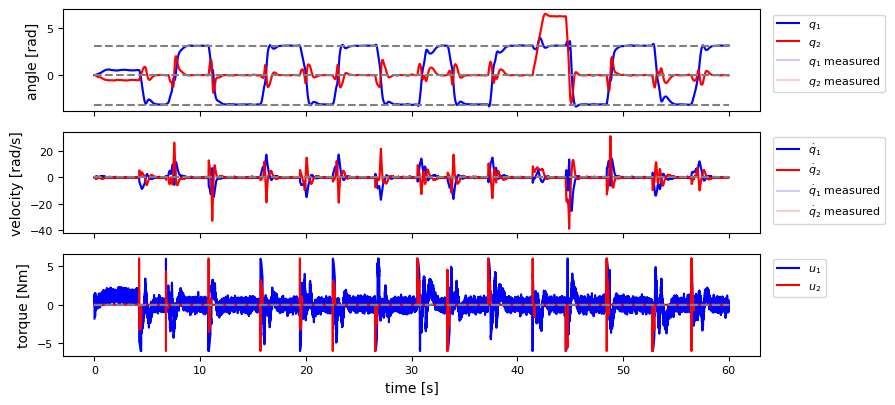}
        \caption{Time series results for VIMPPI-AR-EAPO controller on the pendubot.}
        \label{fig:pendubot_timeseries}

        \vspace{1em}

        \captionof{table}{Pendubot controller performance.}
        \label{tab:pendubot_results}
        \begin{tabular}{lll}
            \toprule
            Controller                         & Swingups         & Uptime [s]        \\
            \midrule
            \rowcolor{LightRed} VIMPPI-AR-EAPO & $16.05 \pm 0.22$ & $48.03 \pm 0.9$   \\
            \rowcolor{LightRed} VIMPPI         & $15.95 \pm 0.22$ & $47.63 \pm 1.12$  \\
            AR-EAPO                            & $17.3 \pm 1.31$  & $39.54 \pm 2.46$  \\
            EMPPI                              & $15.25 \pm 3.74$ & $36.35 \pm 11.33$ \\
            evolsac                            & $24.4 \pm 1.77$  & $31.17 \pm 3.56$  \\
            History SAC                        & $24.2 \pm 3.22$  & $28.07 \pm 1.8$   \\
            IMPPI-AR-EAPO                      & $26.8 \pm 5.62$  & $23.9 \pm 9.68$   \\
            IMPPI                              & $30.65 \pm 6.11$ & $21.61 \pm 9.38$  \\
            IFMPPI                             & $13.5 \pm 7.92$  & $9.21 \pm 6.86$   \\
            mcpilco                            & $11.7 \pm 13.06$ & $4.08 \pm 1.7$    \\
            TVLQR                              & $15.0 \pm 19.92$ & $2.34 \pm 3.45$   \\
            \bottomrule
        \end{tabular}
        \vspace{-1em}
    \end{minipage}
\end{figure}

For the pendubot configuration, where actuation is applied to the first joint, VIMPPI achieved 48 seconds of uptime, outperforming the next-best controller by approximately 8 seconds with half the standard deviation (Table \ref{tab:pendubot_results}). The time series data (Fig. \ref{fig:pendubot_timeseries}) shows VIMPPI maintaining stability even after disturbances. Interestingly, adding a warm-start mechanism primarily reduced variance rather than improving mean uptime, suggesting that the base controller already handles disturbances effectively through its extended planning horizon.

\subsection{Acrobot}

\begin{figure}[t]
    \centering
    \begin{minipage}{\linewidth}
        \centering
        \includegraphics[width=1.0\linewidth]{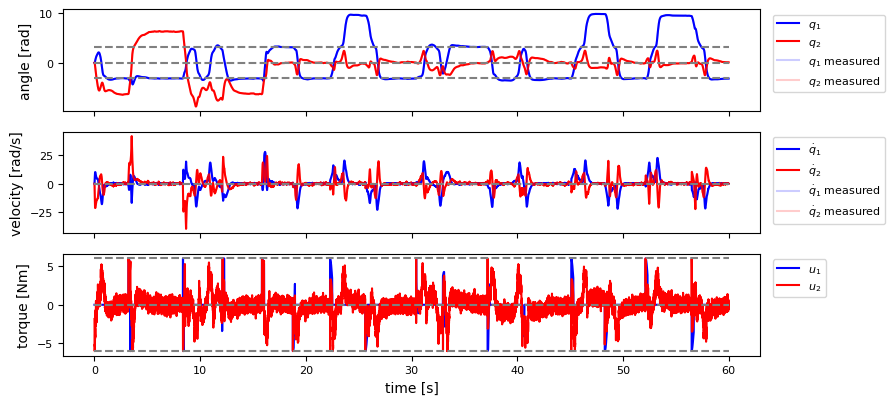}
        \caption{Time series results for VIMPPI-AR-EAPO controller on the acrobot.}
        \label{fig:acrobot_timeseries}

        \vspace{1em}

        \captionof{table}{Acrobot controller performance.}
        \label{tab:acrobot_results}
        \begin{tabular}{lll}
            \toprule
            Controller                         & Swingups          & Uptime [s]       \\
            \midrule
            \rowcolor{LightRed} VIMPPI         & $19.9 \pm 1.76$   & $38.6 \pm 2.51$  \\
            \rowcolor{LightRed} VIMPPI-AR-EAPO & $21.0 \pm 2.76$   & $37.69 \pm 2.3$  \\
            EMPPI                              & $24.0 \pm 2.35$   & $27.24 \pm 3.35$ \\
            IFMPPI                             & $29.3 \pm 2.63$   & $21.96 \pm 3.25$ \\
            AR-EAPO                            & $18.65 \pm 2.46$  & $21.4 \pm 2.12$  \\
            IMPPI-AR-EAPO                      & $32.95 \pm 5.9$   & $19.49 \pm 6.8$  \\
            evolsac                            & $29.0 \pm 4.02$   & $18.79 \pm 3.84$ \\
            IMPPI                              & $34.1 \pm 4.92$   & $17.68 \pm 7.05$ \\
            History SAC                        & $27.55 \pm 4.425$ & $8.9 \pm 1.43$   \\
            TVLQR                              & $7.45 \pm 6.16$   & $1.09 \pm 1.75$  \\
            \bottomrule
        \end{tabular}
        \vspace{-1em}
    \end{minipage}
\end{figure}

The acrobot configuration presents a more challenging control problem due to actuation at the second joint rather than the first, which reduces direct control authority. Nevertheless, the performance gap between VIMPPI and the next-best controller is even more pronounced, reaching approximately 11 seconds (Table \ref{tab:acrobot_results}). This demonstrates that the extended planning horizon is particularly beneficial for systems with limited control authority, where accurate long-term prediction becomes even more critical for successful control.

Interestingly, the warm-started version performed slightly worse than the standard implementation for the acrobot. We hypothesize that AR-EAPO occasionally suggests suboptimal initial trajectories that require additional iterations for MPPI to correct, slightly reducing overall performance. This highlights the importance of carefully considering initialization strategies when dealing with highly dynamic systems.

\section{Conclusion}
Our VIMPPI controller demonstrated significant performance improvements in the AI Olympics competition. By incorporating a variational integrator into the MPPI framework, we achieved a 4-20x increase in effective planning horizon without additional computational cost. This approach consistently outperformed both alternative MPPI implementations and baseline controllers for both the acrobot and pendubot configurations.

The key innovation — replacing standard numerical integration with variational integration in the MPPI rollout — offers an efficient solution for controlling highly dynamic systems without requiring complex reward engineering or system identification. The results highlight the importance of numerical integration methods in sampling-based control approaches, an aspect that has received relatively little attention in the literature.

Future work will focus on applying this technique to more complex robotic systems such as humanoid robots, quadrupeds, and multi-link manipulators, and exploring hybrid approaches that combine our method with learning-based techniques. We believe that the principles demonstrated in this work have broad applicability across robotics and control, particularly for systems where accurate long-term prediction is essential for effective control.

\bibliographystyle{IEEEtran}
\bibliography{ref}
\end{document}